\documentclass{article}
\usepackage{dcase2020,amsmath,graphicx,url,times,booktabs, tabularx}
\usepackage{bm}
\usepackage{amssymb}
\usepackage{cite}
\usepackage{url}
\usepackage{setspace}

\newcommand{\bhline}[1]{\noalign{\hrule height #1}}
   
\title{Description and Discussion on DCASE2020 Challenge Task2: Unsupervised Anomalous Sound Detection for Machine Condition Monitoring}

\name{
Yuma Koizumi$^{1}$,
Yohei Kawaguchi$^{2}$,
Keisuke Imoto$^{3}$,
Toshiki Nakamura$^{2}$,
Yuki Nikaido$^{2}$,
Ryo Tanabe$^{2}$,
}
 \secondlinename{	  
Harsh Purohit$^{2}$,
Kaori Suefusa$^{2}$,
Takashi Endo$^{2}$,
Masahiro Yasuda$^{1}$,
Noboru Harada$^{1}$,
       }
\address{
$^1$ NTT Corporation, Japan, \url{koizumi.yuma@ieee.org}\\           
$^2$ Hitach, Ltd., Japan, \url{yohei.kawaguchi.xk@hitachi.com}\\           
$^3$ Doshisha University, Japan, \url{keisuke.imoto@ieee.org}
  }

\begin{document}

\ninept
\maketitle

\begin{sloppy}

\begin{abstract}
In this paper, we present the task description and discuss the results of the DCASE 2020 Challenge Task 2: Unsupervised Detection of Anomalous Sounds for Machine Condition Monitoring. The goal of anomalous sound detection (ASD) is to identify whether the sound emitted from a target machine is normal or anomalous. The main challenge of this task is to detect unknown anomalous sounds under the condition that only normal sound samples have been provided as training data. We have designed this challenge as the first benchmark of ASD research, which includes a large-scale dataset, evaluation metrics, and a simple baseline system. We received 117 submissions from 40 teams, and several novel approaches have been developed as a result of this challenge. On the basis of the analysis of the evaluation results, we discuss two new approaches and their problems.
\end{abstract}


\section{Introduction}
\label{sec:intro}
Anomalous sound detection (ASD) \cite{koizumi_01,kawaguchi_01,koizumi_02,kawaguchi_02,koizumi_3,suefusa_2020} is the task of identifying whether the sound emitted from a target machine is normal or anomalous. Automatic detection of mechanical failure is an essential technology in the fourth industrial revolution, which includes artificial intelligence (AI)-based factory automation, and also prompt detection of machine anomaly by observing its sounds may be useful for machine condition monitoring. To connect the Detection and Classification of Acoustic Scenes and Events (DCASE) challenge tasks and real-world problems, we organize a new DCASE Challenge task: ``{\it unsupervised ASD}''.

The main challenge of this task is to detect unknown anomalous sounds under the condition that only normal sound samples have been provided as training data \cite{koizumi_01,kawaguchi_01,koizumi_02,kawaguchi_02,koizumi_3,suefusa_2020}. In real-world factories, actual anomalous sounds rarely occur but are highly diverse. Therefore, exhaustive patterns of anomalous sounds are impossible to deliberately make and/or collect. This means that we must detect {\it unknown} anomalous sounds that were not in the given training data. This point is one of the major differences in premise between ASD for industrial equipment and the past supervised DCASE challenge tasks for detecting {\it defined} anomalous sounds such as gunshots or a baby crying \cite{DCASE2017}.

The importance of unsupervised ASD has been recognized for a long time, and various approaches have been investigated \cite{ito,chan,doukas,Ntalampiras_2011,lecomte,conte,bardeli,aurino,kawaguchi_03,giri}.
In early studies, acoustic features for detecting anomalies were designed on the basis of the mechanical structure of the target machine \cite{londono,minemura,wei}. 
Benefiting from the development of deep learning, deep neural network (DNN)-based methods that do not require knowledge of the target machine are also being actively studied \cite{Marchi_2015,Marchi_2015_IJCNN,Oh,Kawachi_2018,Kawachi_2019,adaflow,sniper,spidernet}. 
Although recent studies yielded large-scale datasets for 
ASD \cite{idmt_engine,toyadmos,mimii}, in many DNN-based ASD studies, different dataset and metrics were used, making it difficult to objectively compare the effectiveness and characteristics of these methods.
We believe that providing a benchmark for ASD by designing a unified dataset and metrics will contribute to accelerating both the basic research and industrial use of the latest technologies.

We have designed a DCASE challenge task as the first benchmark of ASD research. The dataset, evaluation metrics, a simple baseline system, and other detailed rules are designed so that they did not deviate from the real-world issues. 
Through the analysis of all 117 submissions, we found that several teams independently developed two new approaches for effectively using the provided normal samples of various types of machine. After briefly introducing this task in Section 3, we discuss the strategies and potential problems of these new methods, and state the next research agenda of unsupervised ASD in Section 4.

\begin{table*}[ttt]
\caption{Baseline system results. Average score and standard deviation from these 10 independent trials.}
\label{tab:result}
\vspace{3pt}
\begin{minipage}{0.33\hsize}
\center
(a) Toy-car
\begin{tabular}{ c| cc }
\bhline{1.3pt}
ID        & AUC [\%]                 & $p$AUC [\%]  \\ \hline
1 (dev.)   & $81.36 \pm 1.15$   & $68.40 \pm 0.92$ \\
2 (dev.)   & $85.97 \pm 0.58$   & $77.72 \pm 0.90$ \\
3 (dev.)   & $63.30 \pm 1.03$   & $55.21 \pm 0.37$ \\
4 (dev.)   & $84.45 \pm 1.87$   & $68.97 \pm 2.37$ \\ \hline
5 (eval.)  & $74.26 \pm 0.63$   & $63.16 \pm 0.75$ \\
6 (eval.)  & $83.38 \pm 1.25$   & $69.92 \pm 0.65$ \\
7 (eval.)  & $82.79 \pm 0.75$   & $65.43 \pm 1.32$ \\
\bhline{1.3pt}
\end{tabular}
\end{minipage}
\begin{minipage}{0.33\hsize}
\center
(b) Toy-conveyor
\begin{tabular}{ c| cc }
\bhline{1.3pt}
ID        & AUC [\%]                 & $p$AUC [\%]  \\ \hline
1 (dev.)   & $78.07 \pm 0.79$   & $64.25 \pm 0.99$ \\
2 (dev.)   & $64.16 \pm 0.53$   & $56.01 \pm 0.71$ \\
3 (dev.)   & $75.35 \pm 1.39$   & $61.03 \pm 1.00$ \\
 &   &  \\ \hline
4 (eval.)  & $92.08 \pm 0.88$   & $72.26 \pm 1.62$ \\
5 (eval.)  & $83.35 \pm 1.75$   & $61.44 \pm 1.56$ \\
6 (eval.)  & $80.66 \pm 1.53$   & $61.15 \pm 2.04$ \\
\bhline{1.3pt}
\end{tabular}
\end{minipage}
\begin{minipage}{0.33\hsize}
\center
(c) Fan
\begin{tabular}{ c| cc }
\bhline{1.3pt}
ID        & AUC [\%]                 & $p$AUC [\%]  \\ \hline
0 (dev.)   & $54.41 \pm 0.47$   & $49.37 \pm 0.10$ \\
2 (dev.)   & $73.40 \pm 0.58$   & $54.81 \pm 0.34$ \\
4 (dev.)   & $61.61 \pm 1.08$   & $53.26 \pm 0.40$ \\
6 (dev.)   & $73.92 \pm 0.54$   & $52.35 \pm 0.51$ \\ \hline
1 (eval.)  & $77.20 \pm 1.23$   & $61.27 \pm 0.78$ \\
3 (eval.)  & $85.61 \pm 0.66$   & $66.50 \pm 0.77$ \\
5 (eval.)  & $85.60 \pm 0.75$   & $69.61 \pm 0.98$ \\
\bhline{1.3pt}
\end{tabular}
\end{minipage}
\\ 
\vspace{3pt}
\\
\begin{minipage}{0.33\hsize}
\center
(d) Pump
\begin{tabular}{ c| cc }
\bhline{1.3pt}
ID        & AUC [\%]                 & $p$AUC [\%]  \\ \hline
0 (dev.)   & $67.15 \pm 0.87$   & $56.74 \pm 0.82$ \\
2 (dev.)   & $61.53 \pm 0.97$   & $58.10 \pm 0.93$ \\
4 (dev.)   & $88.33 \pm 0.66$   & $67.10 \pm 1.09$ \\
6 (dev.)   & $74.55 \pm 1.45$   & $58.02 \pm 1.21$ \\ \hline
1 (eval.)  & $85.26 \pm 0.43$   & $69.53 \pm 0.90$  \\
3 (eval.)  & $79.44 \pm 0.53$   & $60.60 \pm 0.64$  \\
5 (eval.)  & $82.42 \pm 0.11$   & $62.20 \pm 0.78$  \\
\bhline{1.3pt}
\end{tabular}
\end{minipage}
\begin{minipage}{0.33\hsize}
\center
(e) Slide rail
\begin{tabular}{ c| cc }
\bhline{1.3pt}
ID        & AUC [\%]                 & $p$AUC [\%]  \\ \hline
0 (dev.)   & $96.19 \pm 0.43$   & $81.44 \pm 1.89$ \\
2 (dev.)   & $78.97 \pm 0.28$   & $63.68 \pm 0.72$ \\
4 (dev.)   & $94.30 \pm 0.64$   & $71.98 \pm 2.20$ \\
6 (dev.)   & $69.59 \pm 1.45$   & $49.02 \pm 0.41$ \\ \hline
1 (eval.)  & $90.44 \pm 0.91$   & $72.02 \pm 1.39$ \\
3 (eval.)  & $81.96 \pm 0.59$   & $54.75 \pm 0.35$ \\
5 (eval.)  & $65.84 \pm 0.77$   & $50.41 \pm 0.42$ \\
\bhline{1.3pt}
\end{tabular}
\end{minipage}
\begin{minipage}{0.33\hsize}
\center
(f) Valve
\begin{tabular}{ c| cc }
\bhline{1.3pt}
ID        & AUC [\%]                 & $p$AUC [\%]  \\ \hline
0 (dev.)   & $68.76 \pm 0.65$   & $51.70 \pm 0.19$ \\
2 (dev.)   & $68.18 \pm 0.86$   & $51.83 \pm 0.31$ \\
4 (dev.)   & $74.30 \pm 0.71$   & $51.97 \pm 0.20$ \\
6 (dev.)   & $53.90 \pm 0.38$   & $48.43 \pm 0.20$ \\ \hline
1 (eval.)  & $59.33 \pm 1.02$   & $51.75 \pm 0.31$ \\
3 (eval.)  & $56.57 \pm 0.94$   & $51.52 \pm 0.22$ \\
5 (eval.)  & $56.22 \pm 0.65$   & $49.11 \pm 0.12$ \\
\bhline{1.3pt}
\end{tabular}
\end{minipage}
\end{table*}

\section{Unsupervised anomalous sound detection}
\label{sec:preliminaries}

Let $L$-point-long time-domain observation $\bm{x} \in \mathbb{R}^{L}$ be an observation that includes a sound emitted from the target machine.
ASD is an identification problem of determining whether the state of the target machine is normal or anomalous by analyzing $\bm{x}$.

To estimate the state of the target, the anomaly score is calculated; it takes a large value when the input signal seems to be anomalous, and vice versa. To calculate the anomaly score, we construct an anomaly score calculator $\mathcal{A}$ with parameter $\theta$. Then, the target is determined to be anomalous when the anomaly score $\mathcal{A}_{\theta}(\bm{x})$ exceeds the predefined threshold value $\phi$ as
\begin{equation}
\mbox{Decision} = 
\begin{cases}
\mbox{Anomaly}  & (\mathcal{A}_{\theta}(\bm{x}) > \phi) \\
\mbox{Normal}   & \mbox{(otherwise)}\\
\end{cases}.
\label{eq:decision}
\end{equation}

It is obvious from (\ref{eq:decision}) that we need to design $\mathcal{A}$ such that $\mathcal{A}_{\theta}(\bm{x})$ takes a large value when the audio-clip $\bm{x}$ is anomalous. 
Intuitively, this seems to be a design problem of a classifier for a two-class classification problem.
However, it is difficult to solve this task as a simple classification problem, because only normal sound samples are provided as training data in the unsupervised-ASD scenario. 
Thus, the main research question of this task is {\it how can anomalies be detected without actual anomalous training data?}

\section{Task setup}
\subsection{Dataset}
\label{sec:dataset}

The data used for this task comprises parts of ToyADMOS \cite{toyadmos} and the MIMII Dataset \cite{mimii} consisting of the normal/anomalous operating sounds of six types of toy/real machines. Anomalous sounds in these datasets are collected by deliberately damaging the target machines. 
The following six types of toy/real machines are used in this task: Toy-car and Toy-conveyor from ToyADMOS, and Valve, Pump, Fan, and Slide rail from the MIMII Dataset. To simplify the task, we use only the first channel of multichannel recordings; all recordings are regarded as single-channel recordings of a fixed microphone. Each recording is an approximately 10-sec-long audio that includes both the target machine's operating sound and environmental noise. The sampling rate of all signals has been downsampled to 16 kHz. We mixed a target machine sound with environmental noise, and only noisy recordings are provided as training/test data. The environmental noise samples were recorded in several real factory environments. For the details of the recording procedure, please refer to the papers on ToyADMOS \cite{toyadmos} and the MIMII Dataset \cite{mimii}.

In this task, we define two important terms: {\bf Machine Type} and {\bf Machine ID}. Machine Type means the kind of machine, which, in this task, can be one of six: Toy-car, Toy-conveyor, Valve, Pump, Fan, and Slide rail. Machine ID is the identifier of each individual of the same type of machine, which numbers three or four in the training dataset and three in the test dataset.

For each Machine Type and Machine ID, the development dataset includes (i) around 1000 samples of normal sounds for training and (ii) 100--200 samples each of normal and anomalous sounds for the test. The evaluation dataset consists of around 400 test samples each for Machine Type and Machine ID, none of which have a condition label (i.e., normal or anomaly).
Note that the Machine IDs of the evaluation dataset are different from those of the development dataset. Thus, we also provide an additional training dataset that includes around 1,000 normal samples each for Machine Type and Machine ID used in the evaluation dataset.

\subsection{Evaluation metrics}
\label{sec:metric}

This task is evaluated using the area under the receiver operating characteristic (ROC) curve (AUC) and the partial-AUC ($p$AUC). The $p$AUC is an AUC calculated from a portion of the ROC curve over a prespecified range of interest. In our metric, the $p$AUC is calculated as the AUC over a low false-positive-rate (FPR) range $[0,p]$. The AUC and pAUC are respectively defined as
\begin{align}
\mbox{AUC} &= \frac{1}{ N_{-} N_{+} }
\sum_{i=1}^{ N_{-} } \sum_{i=j}^{ N_{+} }
\mathcal{H} \left(
\mathcal{A}_{\theta}(\bm{x}_j^{+})
-
\mathcal{A}_{\theta}(\bm{x}_i^{-})
\right),\\
\mbox{$p$AUC} &= \frac{1}{ \lfloor
pN_{-}
\rfloor N_{+} }
\sum_{i=1}^{ 
\lfloor
pN_{-}
\rfloor
} \sum_{i=j}^{ N_{+} }
\mathcal{H} \left(
\mathcal{A}_{\theta}(\bm{x}_j^{+})
-
\mathcal{A}_{\theta}(\bm{x}_i^{-})
\right),
\end{align}
where $\lfloor \cdot \rfloor$ is the flooring function and $\mathcal{H}(a)$ is the hard-threshold function that returns 1 when $a>0$ and 0 otherwise. 
Here, 
$\{ \bm{x}_i^{-} \}_{i=1}^{N_{-}}$
and
$\{ \bm{x}_j^{+} \}_{j=1}^{N_{+}}$
are normal and anomalous test samples, respectively, and have been sorted so that their anomaly scores are in descending order. 
Here, $N_{-}$ and $N_{+}$ are the numbers of normal and anomalous test samples, respectively.
In this task, we will use $p=0.1$.

The reason for the additional use of the $p$AUC is based on practical requirements. If an ASD system frequently gives false alerts, we cannot trust it, just as ``the boy who cried wolf'' could not be trusted. Therefore, it is especially important to increase the true positive rate (TPR) under low FPR conditions. 
In addition, in the industrial use of an ASD system, stable detection of anomalous sounds in various types of equipment is necessary. Therefore, the overall ranking was determined as the average ranking of all Machine Types; thus, achieving high scores on all Machine Types and IDs is important.

\subsection{Baseline system and results}

\begin{figure*}[t]
  \centering
  \includegraphics[width=175mm,clip]{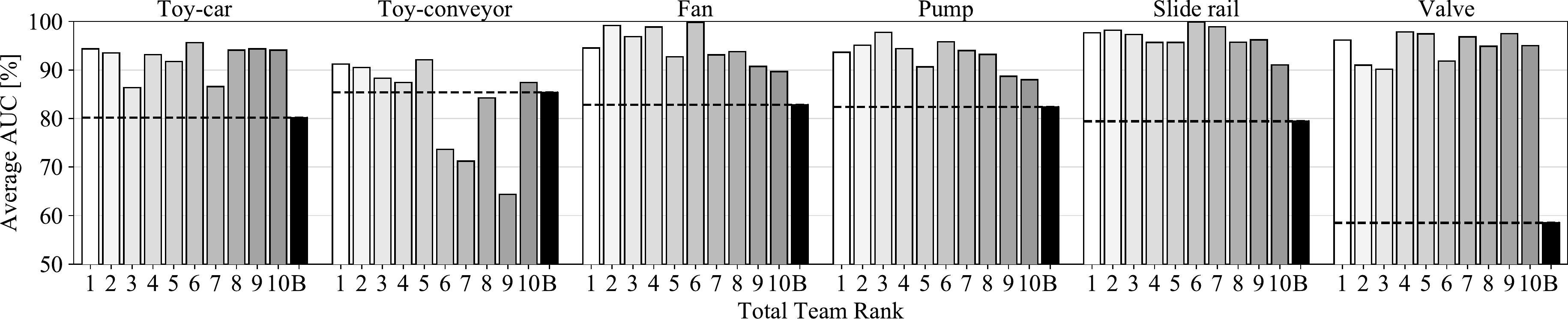}
  \includegraphics[width=175mm,clip]{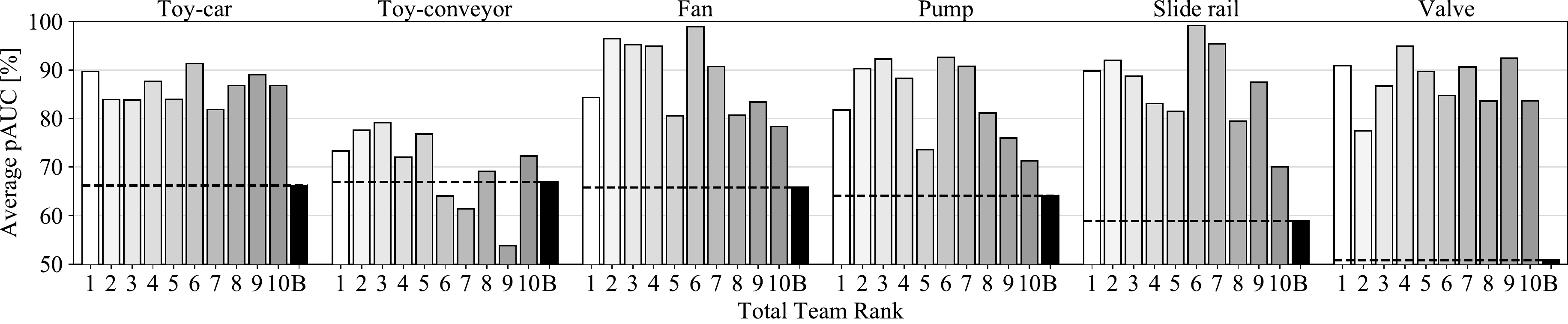} 
  \vspace{-10pt}
  \caption{Evaluation results of top 10 teams in team ranking. (Top) average AUC of each Machine Type, and (bottom) average $p$AUC of each Machine Type. 
  Label ``B'' on the x-axis and black dashed line are baseline system score.}
  \label{fig:result}
  \vspace{-5pt}
\end{figure*}

The baseline system is a simple autoencoder (AE)-based anomaly score calculator. The anomaly score is calculated as the reconstruction error of the observed sound. To obtain small anomaly scores for normal sounds, the AE is trained to minimize the reconstruction error of the normal training data. This method is based on the assumption that the AE cannot reconstruct sounds that are not used in training, that is, unknown anomalous sounds.

In the baseline system, we first calculate a log-mel-spectrogram of the input $\bm{X} \in \mathbb{R}^{ F \times T }$, where $F=64$ and $T$ are the numbers of mel filters and time frames, respectively. 
Then, the log-mel spectrum at $t$ is concatenated with before/after $P$ frames as
$\bm{\phi}_t = ( \bm{X}_{t-P},..., \bm{X}_{t+P})$, using the acoustic feature at $t$.
Then, the anomaly score is calculated as
\begin{align}
\mathcal{A}_{\theta} ( \bm{x} ) = \frac{1}{T} \sum_{t=1}^T 
\lVert \bm{\phi}_t - 
\mathcal{D}_{\theta_D}\left(
\mathcal{E}_{\theta_E}\left( 
\bm{\phi}_t
\right)
\right)
\rVert_2^2,
\end{align}
where 
$\lVert \cdot \rVert _2$ is the $\ell_2$ norm, and 
$\mathcal{E}$ and $\mathcal{D}$ are the encoder and decoder of the AE whose parameters are $\theta_E$ and $\theta_D$, respectively.
Thus, $\theta = \{ \theta_E, \theta_D \}$.
The encoder/decoder of AEs consists of one input fully connected neural-network (FCN) layer, three hidden FCN layers, and one output FCN layer.
Each hidden layer has 128 hidden units, and the dimension of the encoder output is 8. 
The rectified linear unit (ReLU) is used after each FCN layer except the output layer of the decoder.
We stop the training process after 100 epochs, and the batch size is 512.
The ADAM optimizer is used, and we fix the learning rate as 0.001.
We train specialized AEs for each machine Type/ID using only the normal training samples of each Machine Type/ID.

Results obtained with the baseline system are presented in Table\,\ref{tab:result}. Because the results produced with a GPU are generally nondeterministic, the average and standard deviation from these 10 independent trials (training and testing) are shown in the table.

\section{Challenge results}

\subsection{Results for evaluation dataset}
We received 117 submissions from 40 teams and most teams achieved better performance than the baseline system (the team rank of the baseline system was 33rd), which indicates that the challenge was fierce. Because of space limitation, we show the average AUC and $p$AUC of the top 10 teams in the team ranking \cite{e1,e2,e3,e4,e5,e6,e7,e8,e9,e10} in Fig.\,\ref{fig:result}. 
As indicated by the ranking rule of this task, achieving high scores on all Machine Types was important. The top five teams \cite{e1,e2,e3,e4,e5} achieved consistently high scores in all Machine Types. On the other hand, some teams \cite{e6,e7} achieved high scores on several machine types, but, they dropped in the ranks owing to relatively low Toy-conveyor scores. In the following sections, we discuss these two distinctive results.

\subsection{New approach 1: classification-based ASD}

In this task, only around 1,000 samples were provided as training data for each Machine Type/ID, which may be insufficient to train a large-scale DNN. Therefore, it this task, it is important to share the training samples between Machine Type and/or IDs.

We consider one of the novel and interesting approaches for sharing training samples to be ``{\it classification-based ASD}\footnote{Primus called it an {\it outlier-exposed classifier} \cite{e3}, and we might position it as an extension of Fig.\,\ref{fig:img}(b) rather than a classification approach.
}'' \cite{e1,e3,e4,e6,e7}.
Surprisingly, several top-performing teams developed their own classification methods independently.
In the classification approach \cite{e1,e3,e4,e6,e7}, the DNN solves the machine ID identification problem instead of the outlier-detection problem as in the baseline system, as shown in Fig.\,\ref{fig:img}. Eventually, the essence of unsupervised ASD becomes the same as that of classification; finding an accurate decision boundary between normal and others (i.e.,\,anomalies). 
In conventional studies \cite{koizumi_01,koizumi_02,koizumi_3} the AE was trained to increase the anomaly score of simulated anomalous samples using the outlier-detection strategy (Fig.\,\ref{fig:img}(b)). In contrast, many participants adopted the classification strategy, considering the normal samples of other IDs to be anomalies, and they trained the DNN to classify the ID of the input audio (Fig.\,\ref{fig:img}(c)). This strategy enabled both effective use of the data and accurate training of the decision boundary, resulting in a high score in the AUCs of several Machine Types \cite{e4,e6,e7}.

A potential problem with the classification approach might be that it is difficult to train the decision boundaries when notrmal samples of different Machine IDs are very similar. This problem might cause frequent false positives and would be the reason for the lower scores on Toy-conveyor for several teams \cite{e6,e7}. 
To avoid this problem, the winner, Giri et al. \cite{e1}, combined a classification approach and an AE-based approach using the interpolation DNN \cite{suefusa_2020} and achieved stable and accurate detection. On the other hand, Primus \cite{e3} used normal samples of all other Machine IDs and Types as anomalies, and all these other samples were treated as samples in a single anomaly class. Such diversity of anomalies might have contributed to the robust classifier training. We consider that several other possible solutions exist, such as integrating Machine IDs that have similar operating sounds. Thus, classification-based unsupervised ASD may become a future research agenda in unsupervised ASD.

\begin{figure}[t]
  \centering
  \includegraphics[width=85mm,clip]{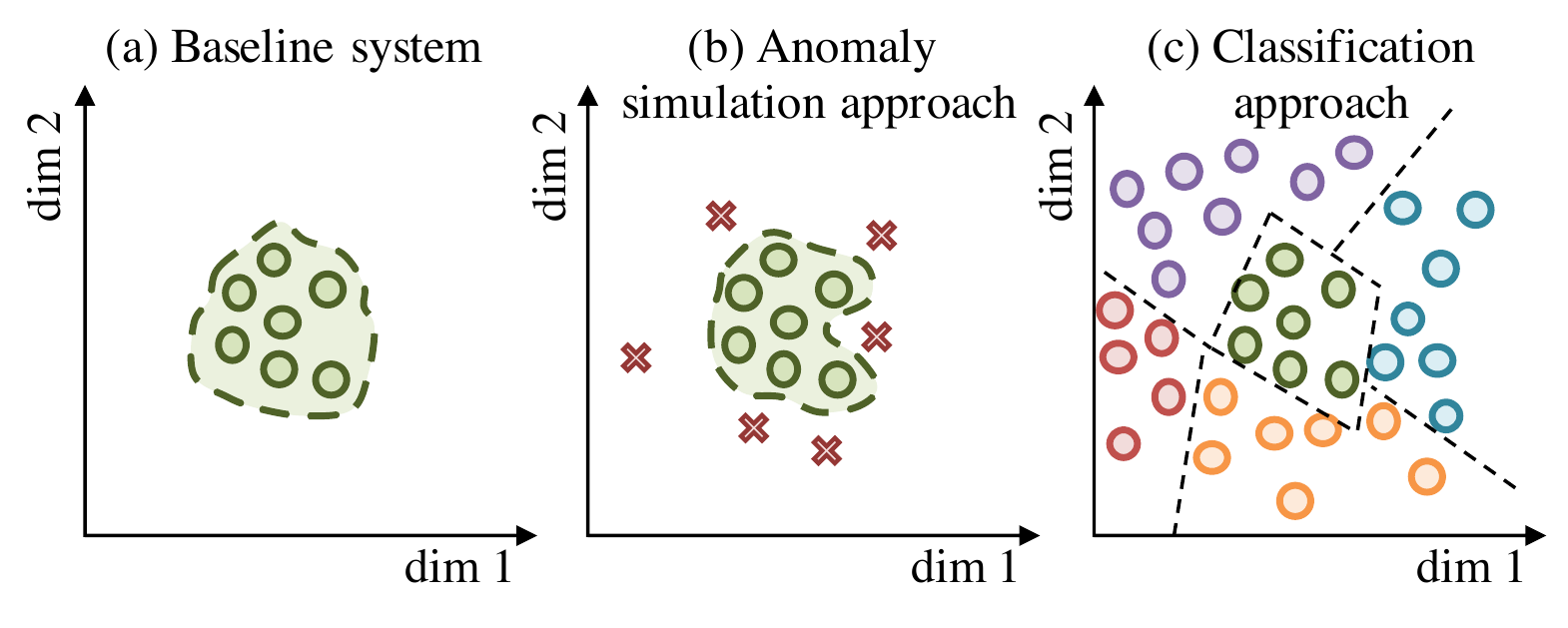}
  \vspace{-15pt}
  \caption{Illustrations of difference between (a)(b) conventional outlier-detection-based approaches and (c) classification approach. Green circles are target Machine IDs normal samples, and dashed lines are decision boundaries. (b) the conventional approach simulates anomalous samples (red crosses) for tightly training decision boundary. In contrast, (c) the classification approach uses other Machine IDs normal samples (other color circles) as anomalous samples for finding the decision boundary.}
  \label{fig:img}
  \vspace{-5pt}
\end{figure}

\subsection{New approach 2: ID conditioning of AE}

Another strategy for sharing training samples is that using Machine ID for the conditioning of AE \cite{e2,e5}, that is, passing the embedded Machine ID to the encoder and/or decoder. 
This approach uses all training samples of the target Machine Type to train AEs while switching the data processing path internally with that of Machine IDs.

The innovative point was to prevent the AE reconstructing various sounds and overlooking anomalous sounds. That is, when the normal sounds of Machine ID (A) are unlike those of Machine ID (B), the AE conditioned by ID (A) must not reconstruct the normal sounds of ID (B) and vice versa.
The 2nd and 5th place teams \cite{e2,e5} developed a method for addressing this problem. 
Daniluk et al. \cite{e2} also input an incorrect label, and then trained the AE to output a predefined constant vector. That is, the AE was trained to be unable to reconstruct the input when the incorrect label is given.
Hayashi et al. \cite{e5} adopted the ID regression approach; they simply concatenated the Machine ID to the input features, and the AE then reconstructed not only the input acoustic features but also the Machine ID. With this approach, we expect that the AE confuses Machine ID when the audio clip includes an anomalous sound.

However, for the same reason as the classification approach, these methods may possibly become a cause of training failure, because when the operating sounds of different Machine IDs are similar, the basic reconstruction error and the additional cost function are antithetical.
From the evaluation results, this problem seems to be not any more critical of a problem than that of the classification approach. However, this problem should also be a future research agenda for the efficient use of training samples.

\section{Conclusions}
\label{sec:concl}

We presented an overview of the task and analysis of the solutions submitted to the DCASE 2020 Challenge Task 2. The main challenge of this task was to detect unknown anomalous sounds under the condition that only normal sound samples have been provided as training data. Several novel approaches were developed as a result of this challenge. We analyzed all evaluation results and submissions, and discussed two new approaches, their problems, and future research directions.

Because of limitations of the space, we could not discuss all innovations, including the identification of a method of data augmentation methods \cite{e4} and a class identification to train an embedding DNN \cite{e5}. We, the organizers, hope that all technical reports of this challenge will be read by many researchers, contributing to advancements in both the academic field and the industrial use of unsupervised ASD.

\bibliographystyle{IEEEtran}
\bibliography{refs}

\begin{thebibliography}{99}
\setlength{\itemsep}{-2.2pt}
\setlength{\baselineskip}{10.5pt}
\vspace{-6pt}
\bibitem{koizumi_01} Y.~Koizumi, S.~Saito, H.~Uematsu, and N.~Harada,
``Optimizing Acoustic Feature Extractor for Anomalous Sound Detection Based on Neyman-Pearson Lemma,''
{\it Proc. of Eur. Signal Process. Conf.} (EUSIPCO), 2017.
\bibitem{kawaguchi_01} Y.~Kawaguchi and T.~Endo,
``How Can We Detect Anomalies from Subsampled Audio Signals?,''
{\it Proc. of IEEE Int'l Workshop on Machine Learning for Signal Process.} (MLSP), 2017.
\bibitem{koizumi_02} Y.~Koizumi, S.~Saito, H.~Uematsu, Y.~Kawachi, and N.~Harada,
``Unsupervised Detection of Anomalous Sound based on Deep Learning and the Neyman-Pearson Lemma,''
{\it IEEE/ACM Trans. on Audio Speech and Language Processing}, 2019.
\bibitem{kawaguchi_02} Y.~Kawaguchi, R.~Tanabe, T.~Endo, K.~Ichige, and K.~Hamada,
``Anomaly Detection Based on an Ensemble of Dereverberation and Anomalous Sound Extraction,''
{\it Proc. of Int'l Conf. on Acous., Speech, and Signal Process.} (ICASSP), 2019.
\bibitem{koizumi_3}
Y.~Koizumi, S.~Saito, M.~Yamaguchi, S.~Murata, and N.~Harada,
``Batch Uniformization for Minimizing Maximum Anomaly Score of DNN-based Anomaly Detection in Sounds,''
\textit{Proc. of the Workshop on Appl. of Signal Process. Audio and Acoust.} (WASPAA), 2019.
\bibitem{suefusa_2020} K.~Suefusa, T.~Nishida, H.~Purohit, R.~Tanabe, T.~Endo, Y.~Kawaguchi,
``Anomalous Sound Detection Based on Interpolation Deep Neural Network,''
{\it Proc. of Int'l Conf. on Acous., Speech, and Signal Process.} (ICASSP), 2020.

\bibitem{DCASE2017} A.~Mesaros, T.~Heittola, A.~Diment, B.~Elizalde, A.~Shah, E.~Vincent, B.~Raj, and T.~Virtanen,
``DCASE 2017 Challenge Setup: Tasks, Datasets and Baseline System,''
{\it Proc. of DCASE}, 2017.

\bibitem{ito} A.~Ito, A.~Aiba, M.~Ito and S.~Makino,
``Detection of Abnormal Sound using Multi-Stage GMM for Surveillance Microphone,''
{\it Proc. Int'l Conf. Info. Assurance and Security}, 2009.
\bibitem{chan} C.~F.~Chan and W.~M.~Eric,
``An Abnormal Sound Detection and Classification System for Surveillance Applications,''
{\it Proc. of Eur. Signal Process. Conf.} (EUSIPCO), 2010.
\bibitem{doukas} C.~N.~Doukas and I.~Maglogiannis,
``Emergency Fall Incidents Detection in Assisted Living Environments Utilizing Motion, sound, and visual perceptual components,''
{\it IEEE Trans. Inf. Technol. Biomed.}, 2011.
\bibitem{Ntalampiras_2011} S.~Ntalampiras, I.~Potamitis, and N.~Fakotakis
``Probabilistic Novelty Detection for Acoustic Surveillance Under Real-World Conditions,''
{\it IEEE Trans. on Multimedia}, 2011.
\bibitem{lecomte} S.~Lecomte, R.~Lengelle, C.~Richard, F.~Capman, and B.~Ravera, ``Abnormal Events Detection using Unsupervised One-Class SVM: Application to audio surveillance and evaluation,''
{\it Proc. of Int'l Conf. Advanced Video and Signal Based Surveillance}, 2011.
\bibitem{conte} D.~Conte, P.~Foggia, G.~Percannella, A.~Saggese and M.~Vento, ``An Ensemble of Rejecting Classifiers for Anomaly Detection of Audio Events,''
{\it Proc. of Int'l Conf. Advanced Video and Signal-Based Surveillance} (AVSS), 2012.
\bibitem{bardeli} R.~Bardeli and D.~Stein, ``Uninformed Abnormal Event Detection on Audio,''
{\it Proc. of ITG Symposium on Speech Communication}, 2012.
\bibitem{aurino} F. Aurino, M. Folla, F. Gargiulo, V. Moscato, A. Picariello, C. Sansone, ``One-Class SVM Based Approach for Detecting Anomalous Audio Events,''
{\it Proc. of Int'l Conf. on Intelligent Networking and Collaborative Systems} (INCoS), 2014. 
\bibitem{kawaguchi_03} Y.~Kawaguchi, T.~Endo, K.~Ichige, and K.~Hamada, ``Non-Negative Novelty Extraction: A New Non-Negativity Constraint for {NMF},” 
{\it Proc. of Int'l Workshop on Acoust. Signal Enh.} (IWAENC), 2018.
\bibitem{giri} R.~Giri, A.~Krishnaswamy, K.~Helwani, ``Robust Non-Negative Block Sparse Coding for Acoustic Novelty Detection,''
{\it Proc. of DCASE}, 2019.

\bibitem{londono} M.~Holgu{\'i}n-Londo{\~n}o, O.~Cardona-Morales, E.~F.~Sierra-Alonso, J.~D.~Mejia-Henao, {\'A}.~Orozco-Guti{\'e}rrez, and G. Castellanos-Dominguez, ``Machine Fault Detection Based on Filter Bank Similarity Features Using Acoustic and Vibration Analysis,''
{\it Math. Probl. Eng.}, 2016.
\bibitem{minemura} K.~Minemura, T.~Ogawa, T.~Kobayashi, ``Acoustic Feature Representation Based on Timbre for Fault Detection of Rotary Machines,''
{\it Proc. of  2018 Int'l Conf. Sensing, Diagnostics, Prognostics, and Control} (SDPC), 2018.
\bibitem{wei} Y. Wei, Y. Li, M. Xu, and W. Huang, ``A Review of Early Fault Diagnosis Approaches and Their Applications in Rotating Machinery,''
{\it Entropy}, vol. 21, no. 4, 2019.


\bibitem{Marchi_2015} E.~Marchi, F.~Vesperini, F.~Eyben, S.~Squartini, and B.~Schuller,
``A Novel Approach for Automatic Acoustic Novelty Detection using a Denoising Autoencoder with Bidirectional LSTM Neural Networks,''
{\it Proc. of Int'l Conf. on Acous., Speech, and Signal Process.} (ICASSP), 2015.
\bibitem{Marchi_2015_IJCNN} E.~Marchi, F.~Vesperini, F.~Weninger, F.~Eyben, S.~Squartini, and B.~Schuller,
``Non-Linear Prediction with LSTM Recurrent Neural Networks for Acoustic Novelty Detection,''
{\it Proc. of Int'l Joint Conf. on Neural Net.} (IJCNN), 2015.
\bibitem{Oh} D.~Y.~Oh, I.~D.~Yun, ``Residual Error Based Anomaly Detection Using Auto-Encoder in {SMD} Machine Sound,''
{\it Sensors}, vol. 18, no. 5, 2018.
\bibitem{Kawachi_2018} Y.~Kawachi, Y.~Koizumi, and N.~Harada,
``Complementary Set Variational Autoencoder for Supervised Anomaly Detection,''
{\it Proc. of Int'l Conf. on Acous., Speech, and Signal Process.} (ICASSP), 2018.
\bibitem{Kawachi_2019} Y.~Kawachi, Y.~Koizumi, S.~Murata and N.~Harada,
``A Two-Class Hyper-Spherical Autoencoder for Supervised Anomaly Detection,''
{\it Proc. of Int'l Conf. on Acous., Speech, and Signal Process.} (ICASSP), 2019.
\bibitem{adaflow} M.~Yamaguchi, Y.~Koizumi, and N.~Harada,
``AdaFlow: Domain-Adaptive Density Estimator with Application to  Anomaly Detection and Unpaired Cross-Domain Transition,''
{\it Proc. of Int'l Conf. on Acous., Speech, and Signal Process.} (ICASSP), 2019.
\bibitem{sniper} Y.~Koizumi, S.~Murata, N.~Harada, S.~Saito, and H.~Uematsu,
``SNIPER: Few-shot Learning for Anomaly Detection to Minimize False-Negative Rate with Ensured True-Positive Rate,''
{\it Proc. of Int'l Conf. on Acous., Speech, and Signal Process.} (ICASSP), 2019.
\bibitem{spidernet} Y.~Koizumi, M.Yasuda, S.~Murata, S.~Saito, and N.~Harada,
``SPIDERnet: Attention Network for One-shot Anomaly Detection in Sounds,''
{\it Proc. of Int'l Conf. on Acous., Speech, and Signal Process.} (ICASSP), 2020.


\bibitem{idmt_engine} 
S.~Grollmisch, J.~Abeßer, J.~Liebetrau, and H.~Lukashevich
``Sounding Industry: Challenges and Datasets for Industrial Sound Analysis'', 
{\it Proc. of Eur. Signal Process. Conf.} (EUSIPCO), 2019.
\bibitem{toyadmos} Y.~Koizumi, S.~Saito, H.~Uematsu, N.~Harada, and K.~Imoto,
``ToyADMOS: A Dataset of Miniature-Machine Operating Sounds for Anomalous Sound Detection,''
\textit{Proc. of the Workshop on Appl. of Signal Process. to Audio and Acoust.} (WASPAA), 2019.
\bibitem{mimii} H.~Purohit, R.~Tanabe, K.~Ichige, T.~Endo, Y.~Nikaido, K.~Suefusa, and Y.~Kawaguchi,
``MIMII Dataset: Sound Dataset for Malfunctioning Industrial Machine Investigation and Inspection,''
\textit{Proc. of DCASE}, 2019.

\bibitem{e1} R.~Giri, S.~V.~Tenneti, F.~Cheng, K.~Helwani, U.~Isik, and A.~Krishnaswamy,
``Unsupervised Anomalous Sound Detection using Self-Supervised Classification and Group Masked Autoencoder for Density Estimation,''
\textit{Tech. report in DCASE2020 Challenge Task 2}, 2020.
\bibitem{e2} 
P.~Daniluk, M.~Go\'zdziewski, S.~Kapka, and M.~Ko\'smider,
``Ensemble of Auto-encoder based and WaveNet like Systems for Unsupervised Anomaly Detection
,''
\textit{Tech. report in DCASE2020 Challenge Task 2}, 2020.
\bibitem{e3} 
P.~Primus,
``Reframing Unsupervised Machine Condition Monitoring as a Supervised Classification Task with Outlier-Exposed Classifiers,''
\textit{Tech. report in DCASE2020 Challenge Task 2}, 2020.
\bibitem{e4} 
T.~Inoue, P.~Vinayavekhin, S.~Morikuni, S.~Wang, T.~H.~Trong, D.~Wood, M.~Tatsubori, and R.~Tachibana,
``Detection of Anomalous Sounds for Machine Condition Monitoring using Classification Confidence,''
\textit{Tech. report in DCASE2020 Challenge Task 2}, 2020.
\bibitem{e5} 
T.~Hayashi, T.~Toshimura, and Y.~Adachi,
``Conformer-based ID-aware Autoencoder for Unsupervised Anomalous Sound Detection
,''
\textit{Tech. report in DCASE2020 Challenge Task 2}, 2020.
\bibitem{e6} 
Q.~Zhou,
``ArcFace based Sound MobileNets for DCASE 2020 Task 2,''
\textit{Tech. report in DCASE2020 Challenge Task 2}, 2020.
\bibitem{e7} 
J.~A.~Lopez, H.~Lu, P.~L-Meyer, L.~Nachman, G.~Stemmer, and J.~Huang,
``A Speaker Recognition Approach to Anomaly Detection,''
\textit{Tech. report in DCASE2020 Challenge Task 2}, 2020.
\bibitem{e8} 
K.~Wilkinghoff, ``Anomalous Sound Detection with Look, Listen, and Learn Embeddings,''
\textit{Tech. report in DCASE2020 Challenge Task 2}, 2020.
\bibitem{e9} 
K.~Durkota, M.~Linda, M.~Ludv\'ik, and J.~Tozicka,
``NEURON-NET: Siamese Network for Anomaly Detection,''
\textit{Tech. report in DCASE2020 Challenge Task 2}, 2020.
\bibitem{e10} 
S.~Grollmisch, D.~Johnson, J.~Abeßer, and H.~Lukashevich,
``IAEO3-Combining OpenL3 Embeddings and Interporation AutoEncoder for Anomalous Sound Detection,''
\textit{Tech. report in DCASE2020 Challenge Task 2}, 2020.

\end{thebibliography}

\end{sloppy}
\end{document}